\documentclass[a4paper]{article} 
\usepackage{graphicx}
\usepackage{amsmath,amssymb}

\begin{document}
\thispagestyle{empty} \parskip=12pt \raggedbottom
 
\def\mytoday#1{{ } \ifcase\month \or January\or February\or March\or
  April\or May\or June\or July\or August\or September\or October\or
  November\or December\fi
  \space \number\year}
\noindent
\vspace*{1cm}
\begin{center}
  {
   \Large  Another weak first order deconfinement transition: three-dimensional
    $SU(5)$ gauge theory} 
\vskip7mm 
 Kieran Holland
\\
\vskip6mm
  Department of Physics,
  University of California San Diego, \\
  9500 Gilman Drive,
  La Jolla CA 92093, USA

\vskip1mm 

  \nopagebreak[4]
 
\begin{abstract}

We examine the finite-temperature deconfinement phase transition of
$(2+1)$-dimensional $SU(5)$ Yang-Mills theory via non-perturbative
lattice simulations. Unsurprisingly, we find that the transition is of
first order, however it appears to be weak. This fits naturally into
the general picture of ``large'' gauge groups having a first order
deconfinement transition, even when the center symmetry associated
with the transition might suggest otherwise.

\end{abstract}
 
\end{center}
\eject

\section{Motivation}

Yang-Mills theory, of self-interacting gluons, is not a full
description of the strong interaction occurring in Nature, but even on
its own it is remarkably rich. At low temperatures, it is a confining
theory, describing a world of color-neutral particles. At high
temperatures, Yang-Mills theory describes weakly-interacting
color-charged deconfined particles in a plasma. This exactly mimics
the behavior of QCD. However, unlike QCD, pure gauge theory has an exact
global symmetry associated with the change from a confining to a
deconfining theory, giving a strict finite-temperature phase
transition. For a more general gauge theory where the number of gluons
is increased e.g.~$SU(N)$, there are hints of a simpler description in terms of
string dynamics in the limit $N \rightarrow \infty$
\cite{'tHooft:1973jz}, a limit which appears surprisingly close to the
real world with $N=3$ \cite{Manohar:1998xv}. There is evidence of
an effective string theory which describes the excitations of the
color flux tube, the QCD string, in the confined phase of the gauge theory 
\cite{Juge:2002br}. Yang-Mills theory is also a testing ground for ideas
about what are the relevant degrees of freedom that lead to
confinement, possible candidates being center vortices, monopoles,
instantons or other topological features
\cite{Engelhardt:2005tu}. From a practical viewpoint, lattice
simulations of pure gauge theory are much less computationally
intensive than those of full QCD, allowing some questions to be
answered in greater detail.  

The global symmetry relevant for the deconfinement transition in pure
gauge theory is the center symmetry
\cite{'tHooft:1977hy}. Finite-temperature $T$ in the gauge theory
means the Euclidean time direction is of finite extent $1/T$ and
physical quantities are periodic in this direction. For gauge group
$G$, the gauge fields themselves are only periodic in the time
direction up to a gauge transformation. The gauge transformations can
be twisted globally by $z \in H$, the center of the group $G$. The
center is the largest subgroup whose elements commute with all
elements of the full group. This is an exact global symmetry of
Yang-Mills theory at finite temperature. At low temperatures, the
theory is confining and the center symmetry is intact. At the critical
temperature $T_c$ where the theory becomes deconfining, the
center symmetry is spontaneously broken. Hence deconfinement in pure
gauge theory is a strict phase transition. Quarks break the center
symmetry explicitly, so the switch from confinement to deconfinement
in QCD is a crossover.   

Over the years, the finite-temperature deconfinement phase transition
in Yang-Mills theory has been studied in great detail and
non-perturbative lattice simulations have played a decisive
role. One can test ideas about the deconfinement transition by varying
the gauge group and the space-time dimensionality. Let us summarize what
is currently known. If the deconfinement transition is of second
order, with a diverging correlation length $\xi$, Svetitsky and Yaffe
conjectured that the universal properties of the transition are
identical to those of the ordering transition of a spin model in one
lower dimension. In particular, the symmetry of the spin system is the center
of the gauge group \cite{Svetitsky:1982gs}. For $SU(N)$, the center is
$Z(N)$, the complex $N$-th roots of 1. In $(3+1)$ dimensions, $SU(2)$
gauge theory does have a second order deconfinement transition
\cite{Kuti:1980gh} and its universal properties, e.g.~how the
correlation length diverges near the critical temperature, 
\begin{equation}
\xi \propto (T-T_c)^{- \nu},
\label{eq:correlation}
\end{equation}
 are identical to those of the 3-dimensional
$Z(2)$-symmetric spin system i.e.~the Ising model
\cite{Engels:1989fz}. Most relevant to Nature, $(3+1)$-dimensional
$SU(3)$ gauge theory has a weak first order deconfinement transition
with a large but finite correlation length \cite{Celik:1983wz}, and
the Svetitsky-Yaffe conjecture does not apply. Continuing this
sequence in $(3+1)$ dimensions, $SU(N)$ gauge theories continue to
have first order deconfinement transitions for $N\ge 4$, with the
transition becoming stronger as $N$ increases \cite{Lucini:2002ku}. In
$(2+1)$ dimensions, the story is somewhat different, with $SU(2)$ and
$SU(3)$ gauge theories both having second order transitions, belonging to the
universality classes of the 2-dimensional $Z(2)$- and $Z(3)$-symmetric spin
models respectively \cite{Teper:1993gp}. For $SU(4)$ the transition
appears very weak but it is not possible to rule out that it is first
order, especially as the $Z(4)$ universality class has a set of
continuously varying critical exponents \cite{deForcrand:2003wa}. 

It is clear that $(3+1)$-dimensional $SU(N)$ gauge theory has a
deconfinement transition that switches from second to
first order as $N$ increases. However, as the center $Z(N)$ also
varies, it's not possible to separate the size of the group from the
change in the center symmetry. An alternate sequence one can consider
is that of the symplectic groups $Sp(N)$, whose center is $Z(2)$ for
all $N$. The group $Sp(N)$ has $N(2N+1)$ generators and there is a
common member $Sp(1)=SU(2)$. Hence one can study the effect of the
size of the group on the deconfinement transition without changing the
symmetry class. What was found is somewhat surprising
\cite{Holland:2003kg}. In $(3+1)$ dimensions, the $Sp(N)$
deconfinement transition changes from second to first order going from
$N=1$ to $N=2$. In $(2+1)$ dimensions, $Sp(N)$ gauge theory has second
order deconfinement transitions for $N=1$ and 2, but it becomes first
order for $N=3$. All second order transitions  belong to the expected
$Z(2)$ universality class. This is the same qualitative behavior as
for $SU(N)$, but in this case the center symmetry is unchanged and one
might {\it a priori} expect the nature of the deconfinement transition
to be the same for all gauge groups.  

A crude argument one can make in favor of first order deconfinement
transitions for large groups is the mismatch of degrees of freedom at
the critical temperature. As the size of the group increases, so does
the number of deconfined gluons in the plasma phase, while the number
of color-neutral states in the confined phase is unchanged. The
results for $Sp(N)$ indicate that the size of the group seems to
dictate the order of the transition, as the $Z(2)$ universality class
is available for all $N$, but the gauge theory chooses not to
avail of it as $N$ increases. The situation is actually similar for
$SU(N)$ in $(3+1)$ dimensions. It turns out that the ordering transitions of
3-dimensional $Z(N)$-symmetric spin models for $N \ge 5$ all belong to
the universality class of the 3-dimensional $U(1)$-symmetric $XY$
model: the $Z(N)$ symmetry is enhanced to $U(1)$ at the critical
point \cite{Hove:2003}. However, the $SU(N)$ gauge theories have first order
deconfinement transitions for $N \ge 3$ and choose not to utilise the
available universality class. A further surprise is given by the exceptional
group $G(2)$, which has 14 generators and whose center is trivially 1.
With a trivial center, there is no distinction between the confined and
deconfined phases and one would expect the deconfinement transition to be a
crossover without any singularity \cite{Holland:2003jy}. In fact, it appears that
$(3+1)$-dimensional $G(2)$ gauge theory actually has an unexpected
first order finite-temperature deconfinement transition
\cite{Pepe:2005}. Again, the size of the group, measured for example
by the number of generators, seems to drive the transition first
order, independent of the symmetry associated with the transition. 

The purpose of this paper is to add one more datum to this collection
of results. We examine $SU(5)$ gauge theory in $(2+1)$ dimensions,
the smallest group which has not yet been studied in this
space-time. Comparing to the known results for $SU(N)$ and $Sp(N)$
gauge theory, one would expect the $SU(5)$ deconfinement transition to
be of first order. The $SU(4)$ transition is very weak and possibly
belongs to the $Z(4)$ universality class, but could also be first
order. The $Sp(3)$ deconfinement transition is weak but clearly of
first order. A first order transition for $SU(5)$ would be consistent
with the general notion that the size of the group dictates the order
of the transition, and we test the idea with this study. There are
additional reasons to expect a first order deconfinement transition
for this theory. The relevant spin model for $SU(5)$ is the
2-dimensional $Z(5)$-symmetric spin model, for which no universality
class is known. In fact, the ordering transition of the $Z(N)$ Potts
model in 2 dimensions is of first order for $N \ge 5$
\cite{Wu:1982ra}. Interestingly, the correlation length of the Potts
model at the critical point behaves as \cite{Klumper:1989} 
\begin{equation}
\xi = \frac{1}{8\sqrt{2}} x (1 + {\cal O}(\frac{1}{x^2})), \hspace{0.75cm}
x = \exp( \frac{\pi^2}{2 \ln \frac{1}{2}(\sqrt{N} + \sqrt{N-4})} ).
\label{eq:xi}
\end{equation}
which diverges as $N \rightarrow 4^+$ and is already on the order of a
few thousand lattice spacings for $N=5$, indicating a very weak first
order transition. This might be an indication that the deconfinement
transition of the gauge theory is first order but weak. In any event,
we wish to rule out any surprises by establishing the nature of the gauge
theory transition using lattice simulations. 

In the course of writing this paper, we learnt of related work by
Liddle and Teper investigating the deconfinement transition of $SU(N)$
gauge theories in $(2+1)$ dimensions for $N=4,5$ and 6
\cite{Teper:2005hx}. We believe they come to the same conclusion that
$SU(5)$ gauge theory has a first order transition. 

\section{Lattice details}

We perform standard lattice simulations of $(2+1)$-dimensional $SU(5)$
gauge theory. We use a finite periodic lattice volume of size $L^2 \times L_t$
and lattice spacing $a$. The fundamental variables are gauge links
$U_{x,\mu}$ which are elements of the group $SU(5)$. We use the
standard Wilson plaquette gauge action
\begin{equation}
S[U] = \frac{10}{g^2} \sum_\Box ( 1 - \frac{1}{5}{\rm Re Tr} U_\Box) = \frac{10}{g^2}
\sum_{x,\mu < \nu} (1 - \frac{1}{5}{\rm Re Tr} (U_{x,\mu} U_{x+\hat{\mu},\nu}
U^{\dagger}_{x+\hat{\nu},\mu} U^{\dagger}_{x,\nu}) ),
\label{eq:action}
\end{equation}
where $g$ is the bare gauge coupling. The partition function is
\begin{equation}
Z = \int {\cal D}U \exp(-S[U]), \hspace{0.75cm} \int {\cal D}U =
\prod_{x,\mu} \int_{SU(5)} dU_{x,\mu}.
\label{eq:partition}
\end{equation}
Finite temperature $T$ in the gauge theory is related to the periodic
Euclidean time extent as $L_t a = 1/T$. The relevant observable to examine the
finite-temperature deconfinement transition is the complex-valued Polyakov loop
\cite{Polyakov:1978vu},
\begin{equation}
\Phi_{\vec{x}} = {\rm Tr} ( {\cal P} \prod_{t=1}^{L_t}
U_{(\vec{x},t),3} ),
\label{eq:polyakov}
\end{equation}
given by a path-ordered product of the time-like gauge links. The
gauge links in the time direction are periodic up to a gauge transformation. The
gauge transformation itself need not be periodic but can be twisted by $z \in Z(5)$.
This center rotation leaves the action $S$ unchanged, hence this is an
exact symmetry of the theory. However the Polyakov loop picks up the twist
$\Phi_{\vec{x}} \rightarrow z \Phi_{\vec{x}}$ because it wraps completely
around the finite time direction, so it is sensitive to the center. The Polyakov loop
expectation value,
\begin{equation}
\langle \Phi \rangle = \frac{1}{Z} \int {\cal D}U \frac{1}{L^2}
\sum_{\vec{x}} \Phi_{\vec{x}} \exp(-S[U]),
\label{eq:vev}
\end{equation}
measures the free energy $F$ at finite temperature
$T$ of a static test quark sitting in the box, $\langle \Phi \rangle =
\exp(-F/T)$. In the confined phase without isolated color-charged
particles, the quark free energy diverges, $\langle \Phi \rangle = 0$ and
the center symmetry is intact, since $\langle \Phi \rangle = 0 = \langle
z \Phi \rangle$. Above the critical temperature $T_c$
where the theory becomes deconfining, color-charged particles in the
plasma have a finite free energy, $\langle \Phi \rangle \ne 0$ and the
center symmetry is spontaneously broken. Hence the deconfinement
transition can be identified via the behavior of the Polyakov loop.

\section{Simulation results}

We use standard methods in our lattice simulations. With the Wilson
gauge action, we generate ensembles of gauge configurations using
heat-bath \cite{Creutz:1980zw} and over-relaxation \cite{Adler:1981sn}
algorithms to update the various $SU(2)$ subgroups of the $SU(5)$
gauge links \cite{Cabibbo:1982zn}. One can update all possible $SU(2)$
subgroups, but this is likely to be an overkill and in practice we
update five randomly chosen subgroups. One sweep of the lattice volume
corresponds to one heat-bath and four over-relaxation updates of every
gauge link.  

\begin{figure}[t]
\begin{center}
\includegraphics[width=0.9\textwidth,height=0.8\textwidth]
{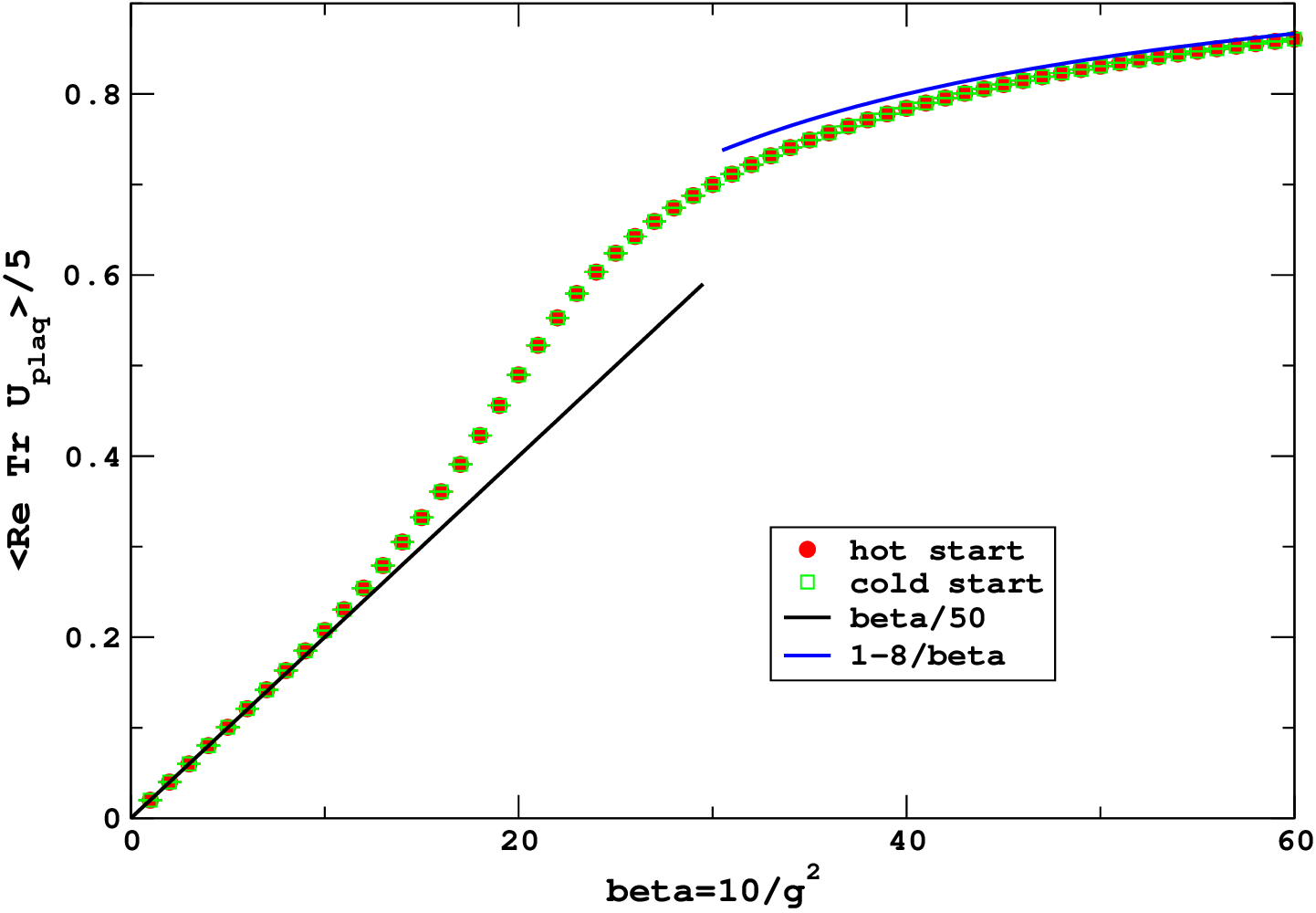}
\end{center}
\vspace{-5mm}
\caption{{}The plaquette expectation value
  $\langle {\rm Re Tr} U_{\Box} \rangle/5$ versus the gauge coupling
  $\beta=10/g^2$. The curves are the lowest order expansions for small
  and large $\beta$. The errors for the data are much smaller than
  the symbol size.}
\label{fig:plaquette}
\end{figure}

The lattice spacing $a$ is implicitly determined by the bare gauge
coupling $\beta=10/g^2$ and the continuum limit is approached by
taking $\beta \rightarrow \infty$. In practice one simulates at a
number of $\beta$ values and tries to extrapolate results to the
continuum. One has to beware of any possible unphysical transitions in
the theory at finite $\beta$. If such a bulk transition exists, this
sets a lower limit on $\beta$ i.e.~an upper limit on the coarsest lattice
spacing one can use and still make a connection to continuum
physics. In Fig.~\ref{fig:plaquette} we plot the expectation value of
the plaquette as a function of the bare gauge coupling on $6^3$
volumes. We use both ordered (``cold'') and random (``hot'') initial
gauge configurations when generating the ensembles, which we see give
completely consistent results. We see no sign of a bulk phase
transition, which would be indicated by a jump in the plaquette average. One
can analytically calculate the plaquette expectation value at weak and strong
coupling, which to leading order give
\begin{eqnarray}
\frac{1}{5} \langle {\rm Re Tr} U_\Box \rangle &=& 1 - 8/\beta 
\hspace{0.5cm} ({\rm large}~\beta) \nonumber \\
&=& \beta/50 \hspace{0.5cm} ({\rm small}~\beta).
\label{eq:strongweak}
\end{eqnarray}
We see that both expansions match excellently the simulation
results. For calibration, we give some of the plaquette expectation
values in Table~\ref{table:plaquette}.

In our simulations to determine the nature of the phase transition, we
consider $L_t=3,4$ and 5 and take the spatial extent as large as
$L=48$. For each $L_t$, there is a critical gauge coupling $\beta_c$
which determines the lattice spacing corresponding to the critical
temperature $T_c=1/L_t a(\beta_c)$. Increasing $L_t$, the lattice
spacing at the critical temperature is reduced and the continuum limit
is approached. In our production runs, for each $\beta$ value and
lattice volume we perform at least 100,000 sweeps to generate the
ensemble of gauge configurations.  

\begin{figure}[t]
\begin{center}
\includegraphics[width=0.95\textwidth,height=0.85\textwidth]
{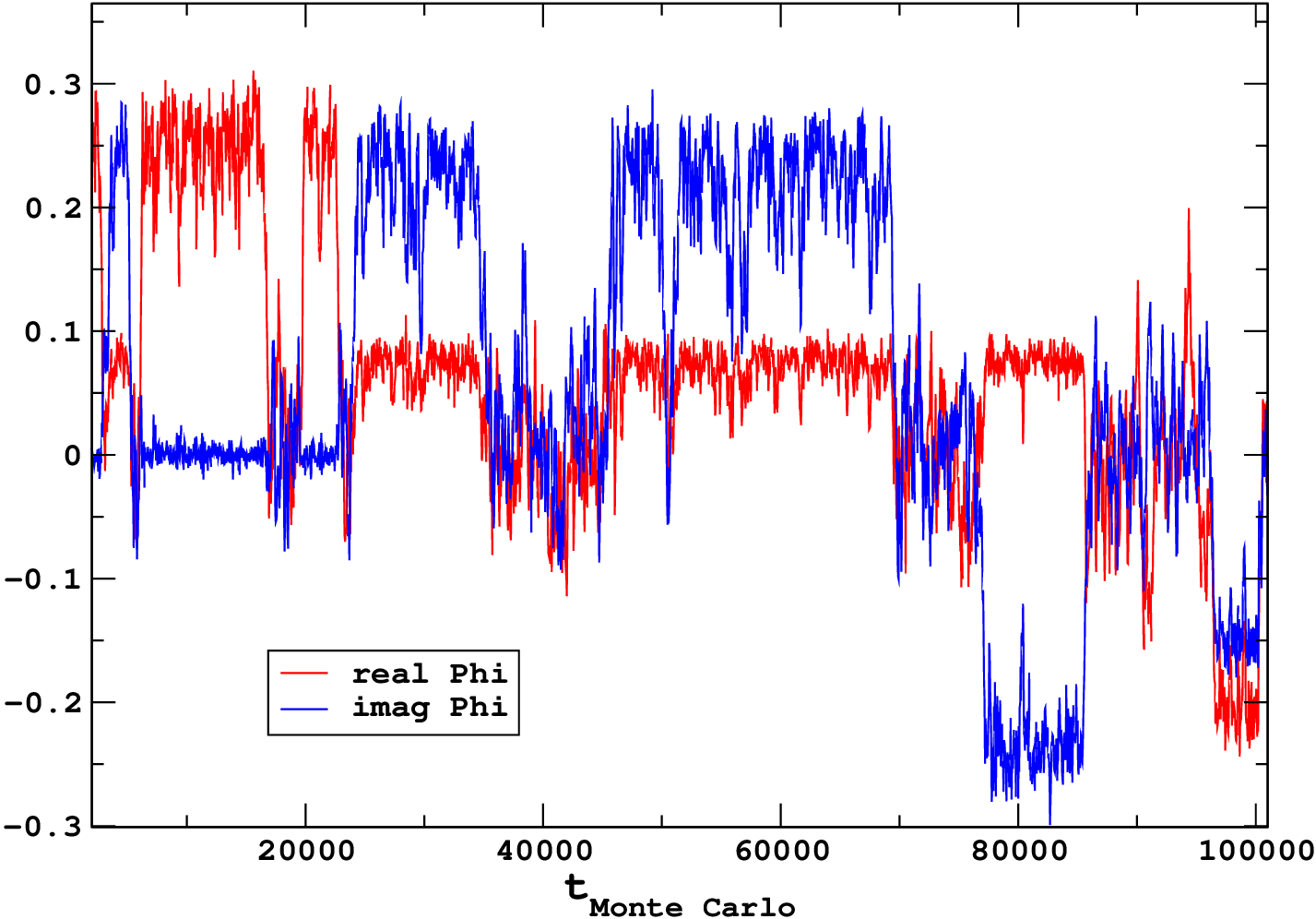}
\end{center}
\vspace{-5mm}
\caption{{}The Monte Carlo history of the complex-valued Polyakov loop
  $\Phi$. The lattice size is $40^2 \times 3$ and the gauge 
  coupling $\beta=32.0513$, which is close to the critical coupling
  for this volume. The system spends long periods in different bulk
  phases separated by rapid tunnelings.}
\label{fig:phi_history}
\end{figure}

The Polyakov loop expectation value $\langle \Phi \rangle$ is the
order parameter which tells us if the system is in the confined or
deconfined bulk phase. In Fig.~\ref{fig:phi_history}, we plot the
Monte Carlo history of the Polyakov loop average configuration by
configuration for a particular lattice size and temperature (the
history is the sequence of gauge configurations generated by the
updating algorithms). The gauge coupling is chosen such that the
temperature is close to criticality. Because of the $Z(5)$ center
symmetry, there are five deconfined bulk phases distinguished by
$\Phi$. We see that the system spends long periods in one bulk phase
before rapidly tunneling to another one. In addition to the periods where
$\Phi\ne 0$, the system spends a considerable fraction of the time
fluctuating around $\Phi \approx 0$.

\begin{figure}[t]
\begin{center}
\includegraphics[width=0.95\textwidth,height=0.85\textwidth]
{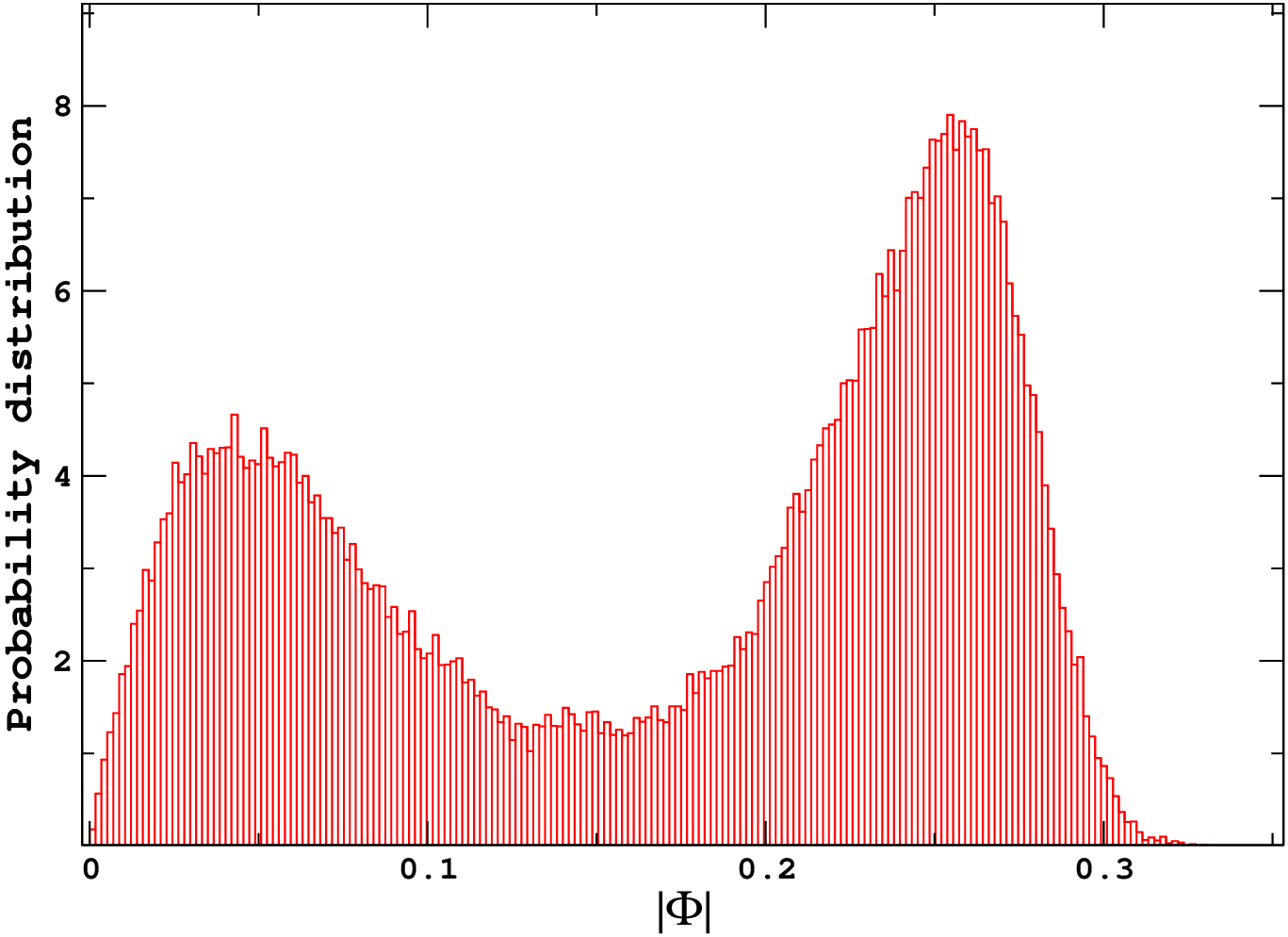}
\end{center}
\vspace{-5mm}
\caption{{}The probability distribution of $|\Phi|$ in
  a volume of size $40^2 \times 3$ and gauge coupling $\beta=32.0513$,
  close to criticality. The double-peaked distribution indicates a
  first order phase transition.}
\label{fig:phi_peaks}
\end{figure}

Because we work at finite volume, the Polyakov loop average over all
gauge configurations will vanish, independently of the temperature, as
the system can tunnel between all possible bulk phases. Only in the
infinite-volume limit is the tunneling suppressed and $\langle \Phi
\rangle \ne 0$. This makes it difficult to locate the critical
temperature where the transition occurs. To eliminate this problem, we
use the modulus $\langle |\Phi| \rangle$ to identify the bulk
phases. This quantity is always non-zero but as the volume increases
$\langle |\Phi| \rangle$ ultimately vanishes in the confined phase and
remains non-zero in the deconfined phase. In
Fig.~\ref{fig:phi_peaks} we plot the probability distribution of
$|\Phi|$ for the ensemble shown in
Fig.~\ref{fig:phi_history}. We see a clear double-peaked distribution,
where we identify the inner and outer peaks with the confined and
deconfined bulk phases respectively. The deconfined phase is slightly
preferred, so the temperature is probably slightly above
criticality. The most important information is that it looks like
there is clearly coexistence of the confined and deconfined phases at
the critical temperature. This is an obvious signal of a first order
deconfinement transition. 

To make the observation more quantitative, we measure the
susceptibility of the Polyakov loop
\begin{equation}
\chi = L^2 [ \langle |\Phi|^2 \rangle - \langle |\Phi| \rangle ^2 ].
\end{equation}
The susceptibility is maximized at a temperature which we use to
define the finite-volume critical coupling $\beta_{c,V}$. This will
differ from other definitions of the critical coupling, but all
methods should agree in the infinite-volume limit. In addition, if the
deconfinement transition is of first order, the rescaled
susceptibility maximum $\chi_{\rm max}/L^2$ should be non-zero as $L
\rightarrow \infty$.

\begin{figure}[t]
\begin{center}
\includegraphics[width=0.95\textwidth,height=0.85\textwidth]
{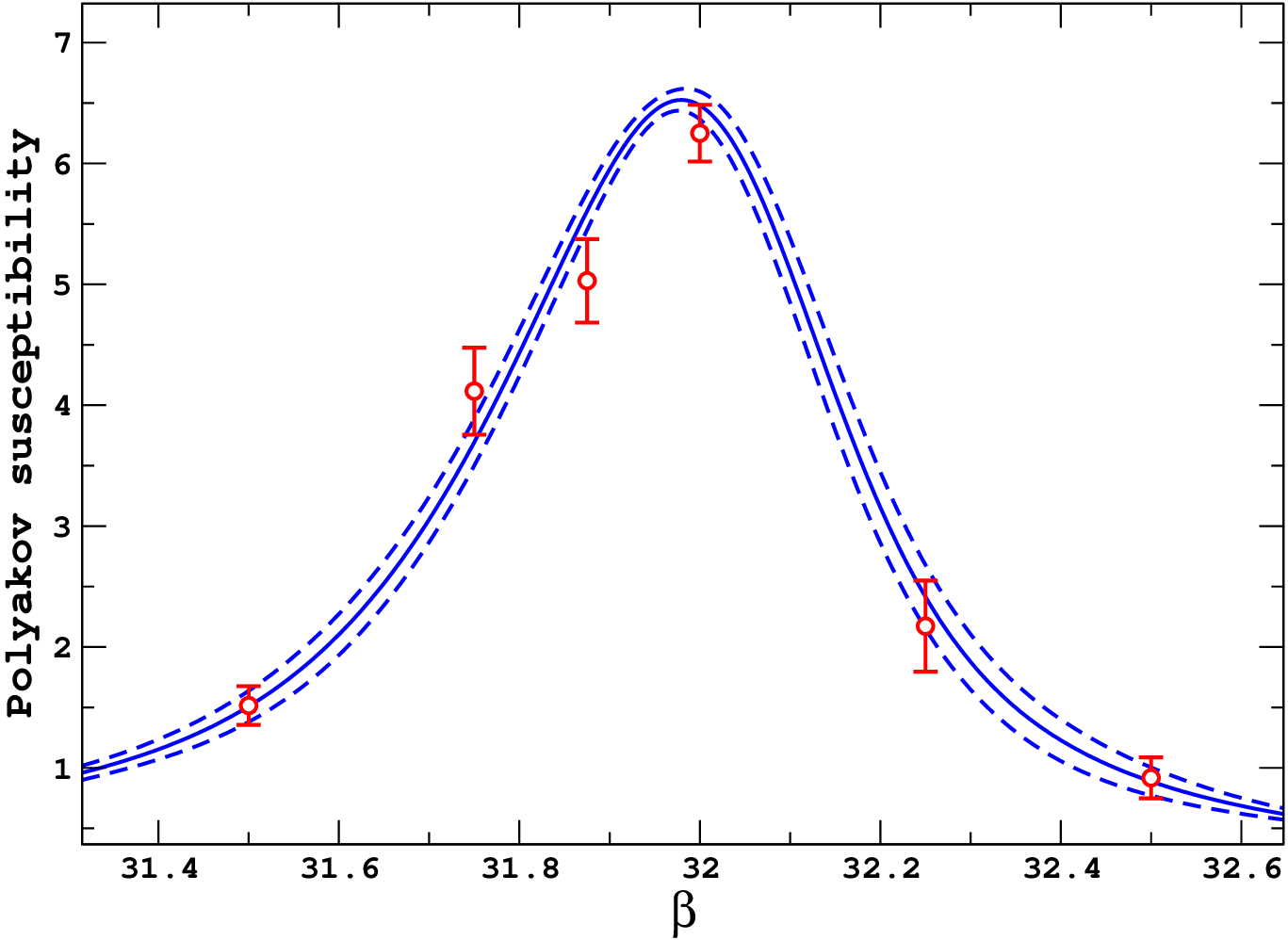}
\end{center}
\vspace{-5mm}
\caption{{}The Polyakov loop susceptibility $\chi$ versus the gauge
  coupling. The data are obtained from separate simulations and the
  curves are obtained using standard reweighting to combine
  the various ensembles. The lattice volume is $28^2 \times 3$.}
\label{fig:chi_reweight}
\end{figure}

For each lattice volume, we perform simulations at a number of gauge
couplings $\beta$. To determine where the susceptibility attains a
maximum, we use the standard reweighting method
\cite{Ferrenberg:1989ui}. A number of ensembles over a range of
$\beta$ values are combined allowing us to interpolate the value of $\chi$ for
intermediary $\beta$ values. This method works excellently provided
there is sufficient overlap among the different ensembles. In
Fig.~\ref{fig:chi_reweight} we plot a typical result using this
method. This allows for an accurate determination of the critical
coupling $\beta_{c,V}$ at the susceptibility peak.

\begin{figure}[t]
\begin{center}
\includegraphics[width=0.95\textwidth,height=0.85\textwidth]
{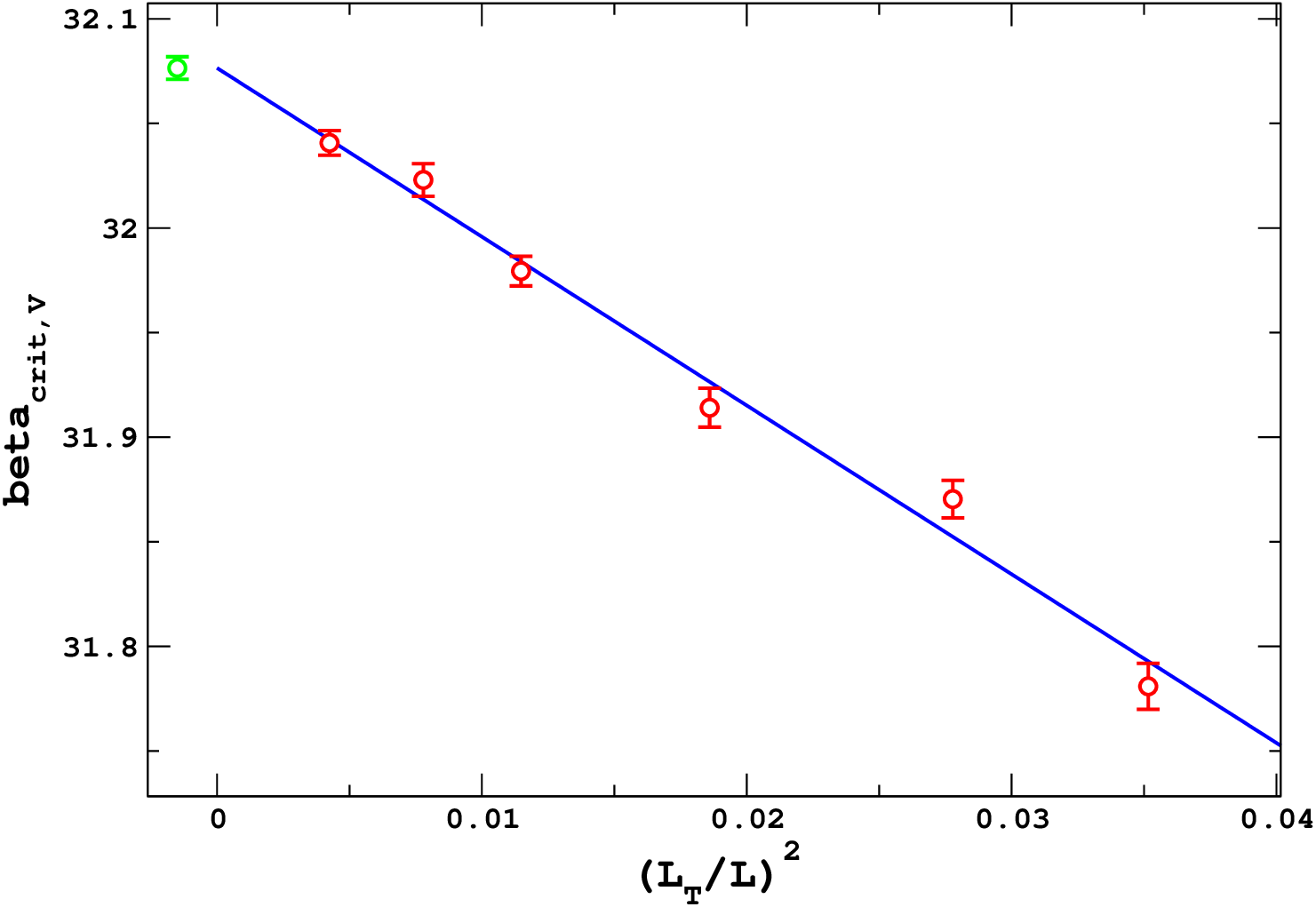}
\end{center}
\vspace{-5mm}
\caption{{}The infinite-volume extrapolation of the critical coupling
  $\beta_{c,V}$ for $L_t=3$, using a $1/L^2$ ansatz. The extrapolated value
  $\beta_{c,\infty}$ is shown slightly offset to the left.}
\label{fig:beta_fit}
\end{figure}

\begin{figure}[t]
\begin{center}
\includegraphics[width=0.95\textwidth,height=0.85\textwidth]
{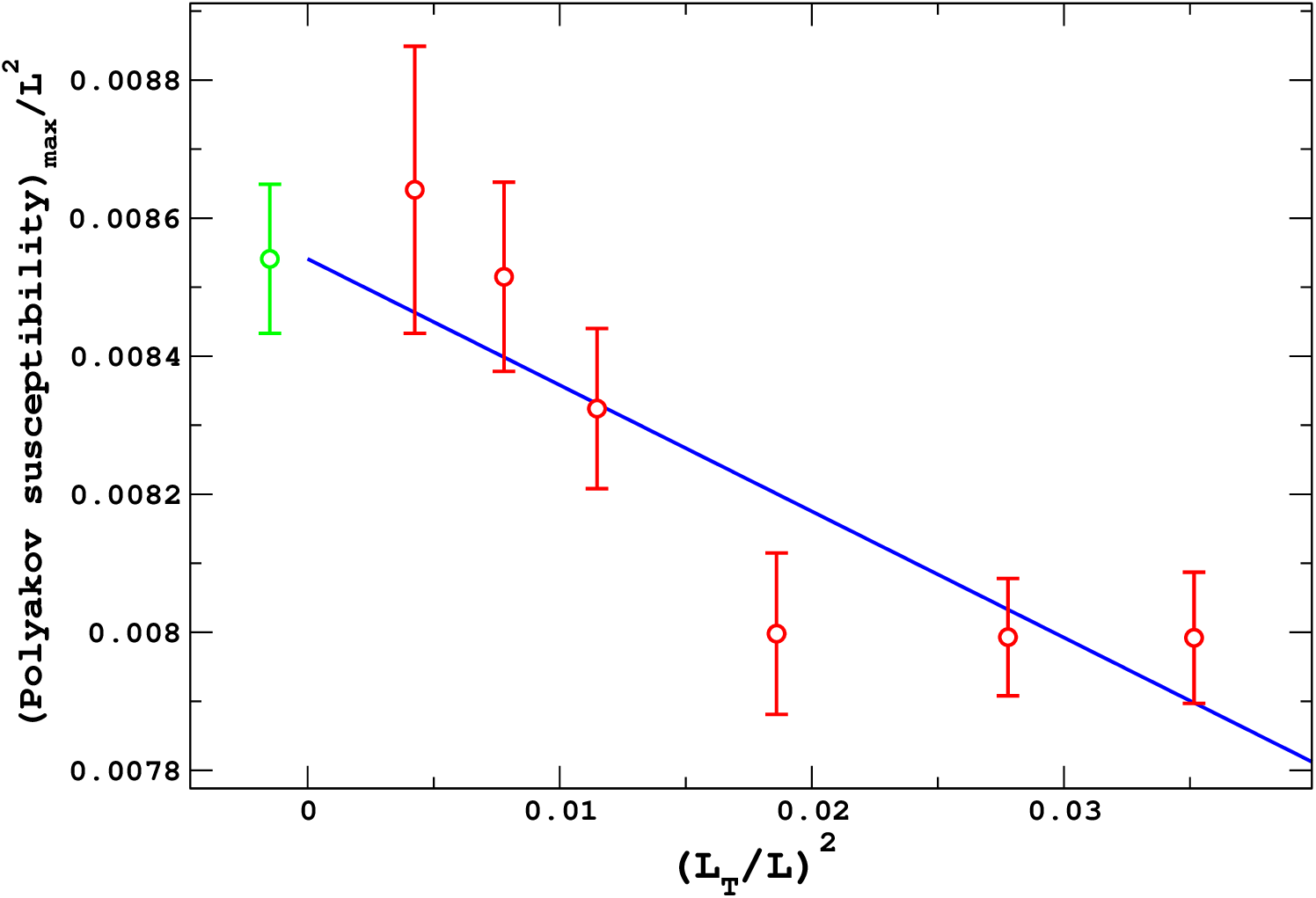}
\end{center}
\vspace{-5mm}
\caption{{}The infinite-volume extrapolation of the susceptibility
  maximum $\chi_{\rm max}/L^2$, using a $1/L^2$ ansatz. The
  extrapolated value $(\chi_{\rm max}/L^2)_{\infty}$ is slightly
  offset to the left.}
\label{fig:chi_fit}
\end{figure}

In Fig.~\ref{fig:beta_fit} we plot the finite-volume critical
couplings $\beta_{c,V}$ determined via reweighting for $L_t=3$. We
extrapolate to the infinite-volume limit using the ansatz
\begin{equation}
\beta_{c,V} = \beta_{c,\infty} + a_0 \frac{L_t^2}{L^2},
\label{eq:beta_fit}
\end{equation}
which we see is in very good agreement with the data. For the
extrapolation of the susceptibility peak, we use the fitting form
\begin{equation}
\frac{\chi_{\rm max}}{L^2} = ( \frac{\chi_{\rm max}}{L^2} )_{\infty} +
b_0 \frac{L_t^2}{L^2}.
\label{eq:chi_fit}
\end{equation}
The data and extrapolation of $\chi_{max}/L^2$ for $L_t=3$ are plotted
in Fig.~\ref{fig:chi_fit}. The fit is reasonable although the data is
not very precisely measured. More importantly, the peak
susceptibility in the infinite-volume limit $(\chi_{\rm
  max}/L^2)_{\infty}$ is clearly non-zero. This is further evidence
that the deconfinement phase transition is of first order.

We find very similar results for $L_t=4$ and 5, namely clear double-peaked
distributions of $|\Phi|$ and a non-zero extrapolated
value for the peak susceptibility $(\chi_{\rm max}/L^2)_{\infty}$. This is good
evidence that the first order deconfinement transition is physical and
remains intact after taking the continuum limit. In Table
\ref{table:extrapolate} we list the extrapolated values for the
critical coupling and peak susceptibility.

The Polyakov loop is an extremely useful observable in determining the
order of the phase transition. In this case, the double-peaked
distribution $P(|\Phi|)$ is a smoking gun that it is first
order. However one would like to see this also reflected in some
thermodynamic quantity. One such observable we call the latent heat
\begin{equation}
\Delta  = \frac{1}{5} ( \langle {\rm Re Tr} U_\Box \rangle_d - \langle
        {\rm Re Tr} U_\Box \rangle_c ),
\label{eq:latent}
\end{equation}
which is given by the difference in the plaquette expectation value
between the confined and deconfined bulk phases. The latent heat is
only defined at the 
critical temperature and strictly speaking only in the infinite-volume
limit. In practice, at finite volume, one can clearly identify using
$\Phi$ which gauge configurations can be classified as being in the
confined or deconfined phase. This identification is only difficult
when the system tunnels from one phase to another, which is a very
small subset of all the gauge configurations and is not a serious
problem. For a first order phase transition, the latent heat is
non-zero, for a second order transition, the latent heat vanishes.

\begin{figure}[t]
\begin{center}
\includegraphics[width=0.95\textwidth,height=0.85\textwidth]
{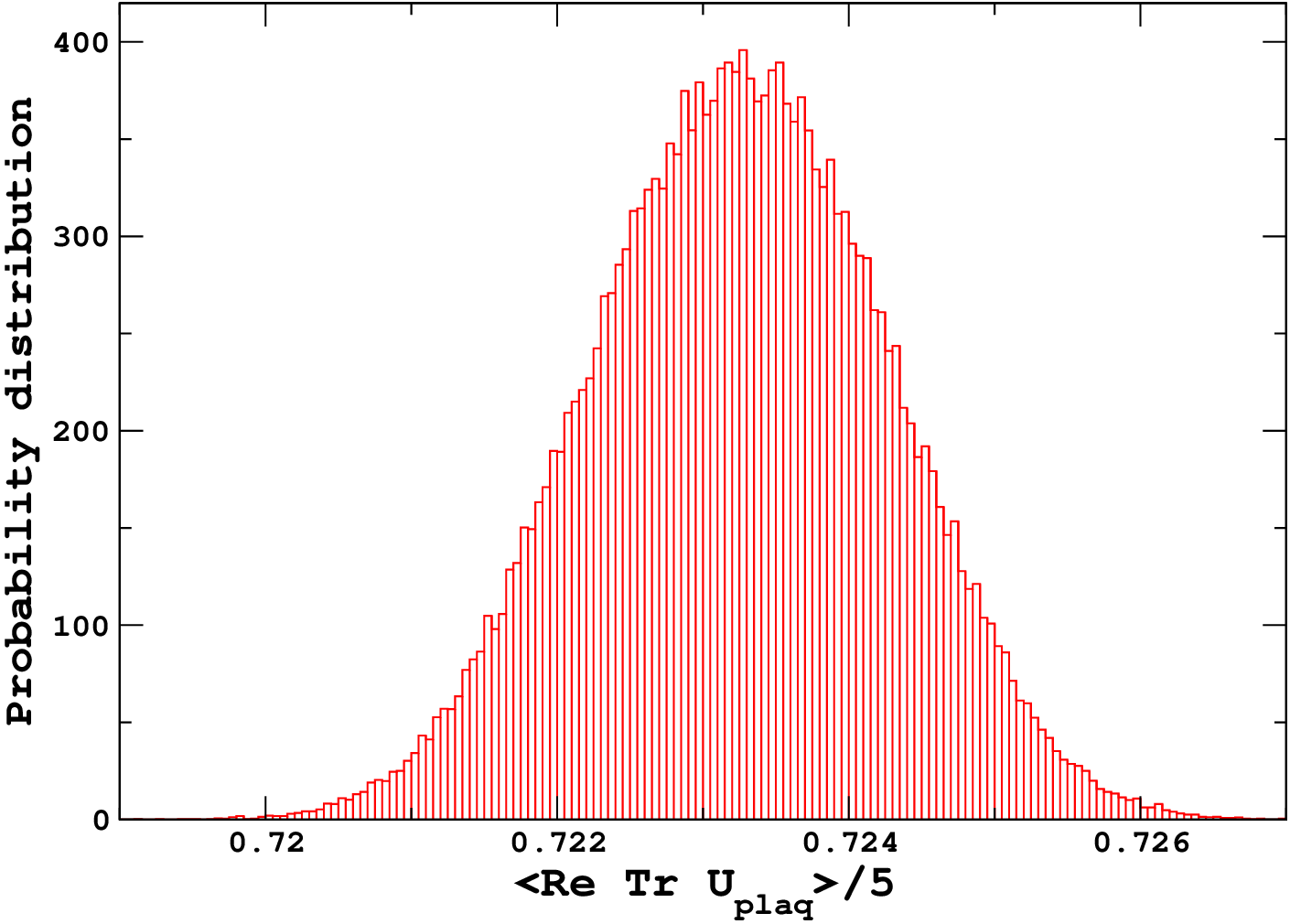}
\end{center}
\vspace{-5mm}
\caption{{}The probability distribution of $\langle {\rm Re Tr} U_\Box
  \rangle/5$ for $SU(5)$ gauge theory in a volume $40^2 \times 3$ at
  gauge coupling $\beta=32.0513$, close to criticality. For a normal
  strength first order transition, one expects two well-separated peaks.}
\label{fig:plaquette_su5}
\end{figure}

\begin{figure}[t]
\begin{center}
\includegraphics[width=0.95\textwidth,height=0.85\textwidth]
{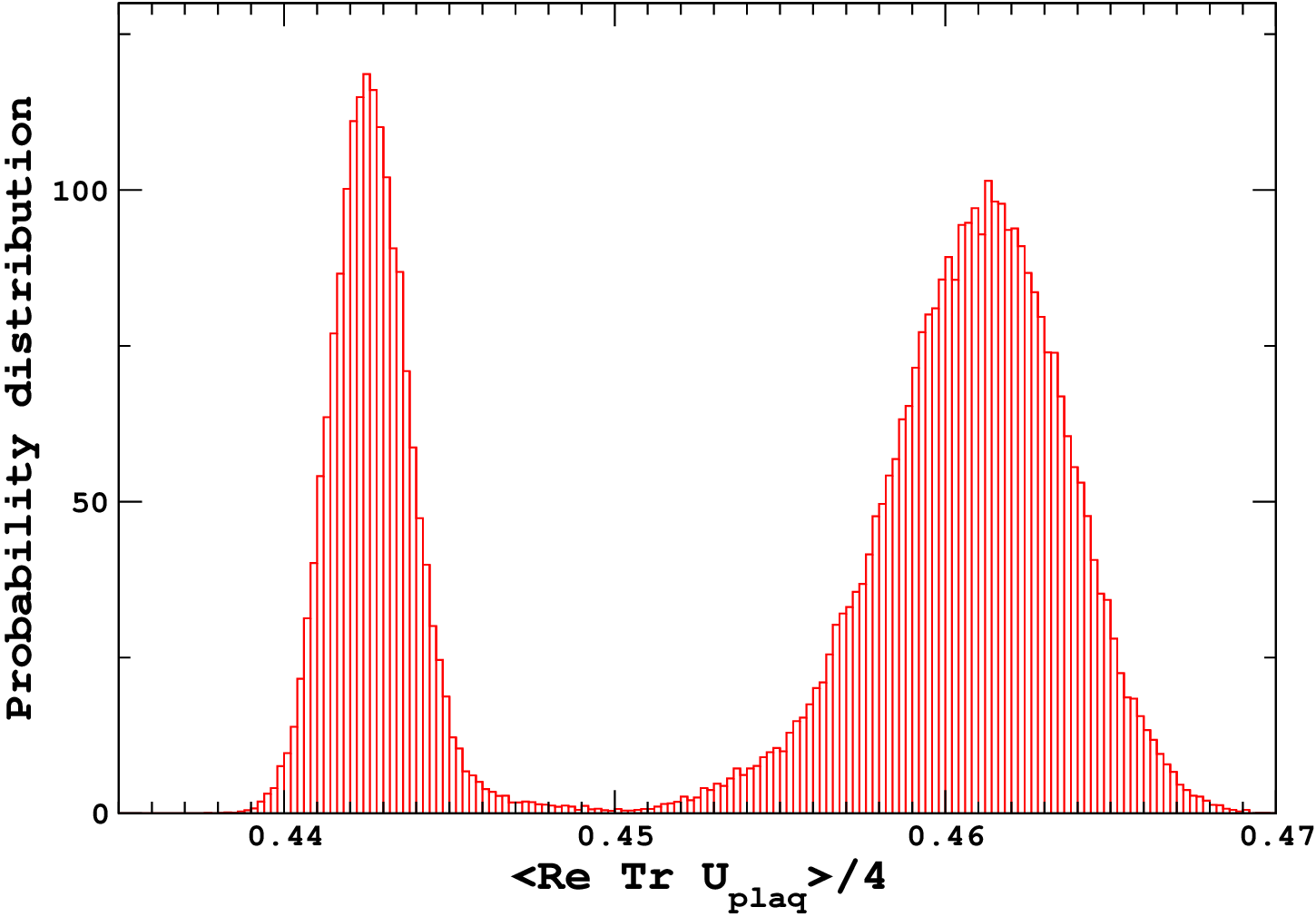}
\end{center}
\vspace{-5mm}
\caption{{}The probability distribution of $\langle {\rm Re Tr} U_\Box
  \rangle/4$ for $Sp(2)$ gauge theory in a volume $20^3 \times 2$ at
  gauge coupling $\beta=6.465$, close to criticality. There are two
  clearly indentifiable peaks, whose separation indicates a normal
  strength first order deconfinement transition.}
\label{fig:plaquette_sp2}
\end{figure}

In Fig.~\ref{fig:plaquette_su5} we plot the distribution of the
plaquette average for an ensemble of gauge configurations in a large
volume very close to the critical temperature for $L_t=3$. For a first
order transition of typical strength, we would expect to see a
double-peaked distribution whose splitting is the latent
heat. Dimensionally we expect $\Delta \sim T_c^4$ so in lattice
spacing units $\Delta a^4 \sim 1/L_t^4$, which on this lattice is
$\approx 0.012$. Since the distribution shows no such splitting, it is
clear that the latent heat is smaller than expected. We note that
the latent heat becomes much more difficult to measure as we approach
the continuum limit, hence we try to extract it at the coarsest
lattice spacing we have used. For comparison, we plot in
Fig.~\ref{fig:plaquette_sp2} the plaquette distribution in
$(3+1)$-dimensional $Sp(2)$ gauge theory in a $20^3 \times 2$ volume,
close to the first order deconfinement transition that occurs in this
theory. We see very distinct peaks with a separation on the order of
the expected $1/L_t^4$. In this case, the latent heat is clearly
non-zero and the first order transition is of typical strength. 

Although we only see a single peak in the plaquette distribution, we
can still extract the latent heat $\Delta$ relatively accurately. Each
configuration can be identified as being in the confined or deconfined
phase using $\Phi$. We separate each ensemble into confined and
deconfined subsets and measure the plaquette expectation value in
each. For a small number of configurations, there is an ambiguity as to which
subset they belong to, an uncertainty we estimate by varying the cut
used to separate the deconfined and confined ensembles. In Table \ref{table:latent}
we list the measured latent heat for $L_t=3$ in a number of
volumes. We see that there is some volume dependence. Using the value
in the largest volume as our best estimate of the infinite-volume
latent heat, we obtain $\Delta/T_c^4 = 0.0867(41)$. The
deconfinement transition is indeed first order, but by this measure
is somewhat weak.

\section{Discussion}

As we stated at the outset, a first order deconfinement transition was
expected in $(2+1)$-dimensional $SU(5)$ gauge theory and we found no
surprises. The fact that the transition appears to be weak, as
measured by the latent heat, might even have been suggested by the
very large correlation length $\xi$ in the 2-dimensional $Z(5)$ Potts
model at the critical point. However, the effective action for the
Polyakov loop has a complicated non-local form. That it shares the
dimensionality and global symmetry of the Potts model does not dictate
its functional form and ultimately the behavior of the deconfinement
transition. It was very unlikely to discover a new universality class
for the phase transition of the gauge theory, but since lattice
simulations can give a definitive answer, we believe we have ruled out
this possibility.

Although is seems clear that the first order transition
of $SU(N)$ gauge theory increases in strength with $N$, it has been
suggested that the $N \rightarrow \infty$ transition might be of
second order \cite{Pisarski:1997yh}. Then the weak first order
transition for $N=3$ in $(3+1)$ dimensions would be a small
perturbation around the large-$N$ limit, consistent with everything
else known about QCD phenomenology using the $1/N$ expansion. This
might be an attractive scenario, but it is completely opposed by all
the evidence.

The general picture appears consistent: where universality classes
are available for ``small'' gauge groups, the deconfinement transition
is of second order and has the universal properties of the ordering
transition of the respective spin model. However the transition
switches to being first order as the gauge group increases in size,
both in $(2+1)$ and $(3+1)$ dimensions, even though there are
available universality classes. It is even more surprising that
$(3+1)$-dimensional $G(2)$ gauge theory has a first order transition,
given that the center is trivial. It is easy to correlate this general behavior
with the size of the gauge groups and speculate that the large number
of degrees of freedom is the driving force. One hopes that this
collection of information can give some insight into the full behavior
of the Polyakov loop effective action or other properties of the gauge theory. 

\section{Acknowledgements}

We would like to thank Michele Pepe for very helpful discussions and
Urs Wenger for the use of his invaluable code to perform the
reweighting of the ensembles. This work was supported by the
U.S.~Department of Energy under the grant DOE-FG03-97ER40546. 



\begin{table}[t]
\begin{center}
\renewcommand{\arraystretch}{1.2} 
\begin{tabular}{cc} 
\hline \hline
$\beta$ & $\langle {\rm Re Tr} U_\Box \rangle/5$ \\
\hline \hline
2.0 & 0.04001(3) \\
4.0 & 0.08020(3) \\
6.0 & 0.12104(3) \\
8.0 & 0.16318(3) \\
10.0 & 0.20737(3) \\
12.0 & 0.25432(4) \\
14.0 & 0.30504(4) \\
16.0 & 0.36066(6) \\
18.0 & 0.42270(7) \\
20.0 & 0.48971(8) \\
22.0 & 0.55268(7) \\
24.0 & 0.60331(6) \\
26.0 & 0.64261(5) \\
28.0 & 0.67409(5) \\
30.0 & 0.70005(4) \\
32.0 & 0.72189(4) \\
34.0 & 0.74076(4) \\
36.0 & 0.75704(4) \\
38.0 & 0.77142(3) \\
40.0 & 0.78411(2) \\
42.0 & 0.79546(2) \\
44.0 & 0.80564(2) \\
46.0 & 0.81478(2) \\
48.0 & 0.82313(2) \\
50.0 & 0.83074(2) \\
52.0 & 0.83773(2) \\
54.0 & 0.84409(2) \\
56.0 & 0.85002(2) \\
58.0 & 0.85549(2) \\
60.0 & 0.86057(2) \\
\hline
\end{tabular}
\end{center}
\caption{{} The plaquette expectation values $\langle {\rm Re Tr} U_\Box
  \rangle/5$ as a function of the bare gauge coupling
  $\beta=10/g^2$ in $6^3$ volumes. The errors are in parentheses.}
\label{table:plaquette}
\end{table}

\begin{table}[t]
\begin{center}
\renewcommand{\arraystretch}{1.2} 
\begin{tabular}{ccc} 
\hline \hline
$L_t$ & $\beta_{c,\infty}$ & $(\chi_{\rm max}/L^2)_{\infty}$ \\
\hline \hline
3 & 32.0765(54) & 0.00854(11) \\
4 & 41.113(12) & 0.00685(13) \\
5 & 50.275(20) & 0.00641(25) \\
\hline
\end{tabular}
\end{center}
\caption{{} The critical gauge coupling and rescaled peak
  susceptibility for $L_t=3,4$ and 5, extrapolated to the infinite-volume
  limit. The estimated errors are in parentheses.}
\label{table:extrapolate}
\end{table}

\begin{table}[t]
\begin{center}
\renewcommand{\arraystretch}{1.2} 
\begin{tabular}{cc} 
\hline \hline
$L$ & $\Delta$ \\
\hline \hline
16 & 0.00130(10) \\
18 & 0.00119(8) \\
22 & 0.00118(5) \\
28 & 0.00113(6) \\
34 & 0.00108(8) \\
40 & 0.00107(5) \\
\hline
\end{tabular}
\end{center}
\caption{{} The latent heat $\Delta$ at the critical
  temperature for $L_t=3$. The estimated errors, in parentheses,
  include the ambiguity in identifying which gauge configurations are
  in the confined and deconfined phase.} 
\label{table:latent}
\end{table}

\eject

\end{document}